\documentclass[prd,twocolumn,floatfix,amsmath,amssymb,floatfix]{revtex4}
\usepackage{graphicx,color,dcolumn,booktabs,bm,mathrsfs}
\usepackage{longtable,lscape}
\usepackage{txfonts}
\usepackage{overpic}
\usepackage{amssymb}
\usepackage{indentfirst}
\usepackage{feynmf}   
\usepackage{slashed}  
\usepackage{cases}
\usepackage{color}
\usepackage{multirow}
\usepackage{epstopdf}
\definecolor{cover}{rgb}{0.77,0.87,0.88}
\definecolor{blueone}{rgb}{0.1,0.1,.7}
\definecolor{citec}{rgb}{0.14,0.47,0.09}
\definecolor{two}{rgb}{0.0,0.5,0.}
\definecolor{three}{rgb}{.5,.1,0.15}
\usepackage[bookmarks=true,
            bookmarksopen=false,
            plainpages=false,
            breaklinks=true,
            colorlinks,
            bookmarksnumbered=true,
            hypertexnames=false,
            filecolor=blue,
            urlcolor=three,
            menucolor=three,
            linkcolor=three,
            citecolor=blueone,
            colorlinks,
            anchorcolor=blue,
            runcolor=pink,
            frenchlinks=red
            pdfstartview=FitH,
            pdftitle=title]{hyperref}
   
\definecolor{cover}{rgb}{0.77,0.87,0.88}
\definecolor{blueone}{rgb}{0.1,0.1,.7}
\definecolor{citec}{rgb}{0.14,0.47,0.09}
\definecolor{two}{rgb}{0.0,0.5,0.}
\definecolor{three}{rgb}{.5,.1,0.15}

\usepackage{relsize}
\usepackage{xspace}
\def\babar{\mbox{\slshape B\kern-0.1em{\smaller A}\kern-0.1em
    B\kern-0.1em{\smaller A\kern-0.2em R}}}

\begin{document}
\title{Molecular picture for $X_{0}(2900)$ and $X_1(2900)$}

\author{Jun He $^1$}\email{junhe@njnu.edu.cn}
\author{Dian-Yong Chen $^2$\footnote{Corresponding author}} \email{chendy@seu.edu.cn}
\affiliation{$^1$Department of  Physics and Institute of Theoretical Physics, Nanjing Normal University, Nanjing 210097, People's Republic of China\\
$^2$School of Physics, Southeast University, Nanjing 210094, People's Republic of China}
\date{\today}

\date{\today}
\begin{abstract}
Inspired by the newly observed $X_{0}(2900)$ and $X_1(2900)$ at LHCb, the $K^*\bar{D}^*$ and $K\bar{D}_1$ interactions are studied in the quasipotential Bethe-Salpeter equation approach combined with the one-boson-exchange model. The bound and virtual states from the interactions are searched for as the poles in the complex energy plane of scattering amplitude.  A  bound state with $I(J^P)=0(0^+)$ and a  virtual state with $0(1^-)$ are produced from the $K^*\bar{D}^*$ interaction and $K\bar{D}_1$ interaction, and can be related to the $X_{0}(2900)$ and $X_1(2900)$ observed at LHCb, respectively. A bound state with $I(J^P)=0(1^+)$ and a virtual state with $I(J^P)=0(2^+)$ are also predicted from the $K^*\bar{D}^*$ interaction with a small $\alpha$ value, which can be searched in future experiments.  	
\end{abstract}

\maketitle
\section{INTRODUCTION}

In the past decades, a growing number of new hadron states have been observed experimentally, then the investigations on the nature of these new hadron states have become one of intriguing topic in hadron physics. Among these new hadron states, some are hardly assigned as conventional mesons or baryons, thus  they were considered as good candidates of  QCD exotic states, such as hadronic molecular states, compact multiquark states and hybrid states (for recent reviews, we refer to Refs.~\cite{Chen:2016qju,Hosaka:2016pey,Richard:2016eis,Lebed:2016hpi,Ali:2017jda,Esposito:2016noz,Guo:2017jvc,Olsen:2017bmm,Karliner:2017qhf,Liu:2019zoy,Brambilla:2019esw}).

Very recently, the LHCb collaboration observed two new states, $X_0(2900)$ and $X_1(2900)$, in the $K^+D^- $ invariant mass distribution of $B^+ \to D^+ D^- K^+$, the resonance parameters of these two states were reported to be \cite{LHCb:2020new},
\begin{eqnarray}
m_{X_0(2900)} &=& (2866\pm 7) \ \mathrm{MeV}, \nonumber\\ 
\Gamma_{X_0(2900)} &=&(57.2 \pm 12.9) \ \mathrm{MeV}, \nonumber\\
m_{X_1(2900)} &=& (2904\pm 5) \ \mathrm{MeV}, \nonumber\\
\Gamma_{X_1(2900)} &=& (110.3 \pm 11.5) \ \mathrm{MeV}, 
\end{eqnarray}  
respectively. The $J^P$ quantum numbers of $X_0(2900)$ and $X_1(2900)$ are $0^+$ and $1^-$, respectively \cite{LHCb:2020new}. 

Since $X_0(2900)$ and $X_1(2900)$ are observed in the $K^+D^- $ channel, the only possible quark components of these states are $u d \bar{c} \bar{s}$, which indicates that they are composed of quarks with four different flavors. Such kind of states are particularly interesting since they obviously can not be assigned as a conventional hadron. Actually, in 2016 another similar structure $X(5568)$ was reported by the D0 collaboration in the $B_s \pi$ invariant mass distribution, which is also a fully open flavor state~\cite{D0:2016mwd}. However, after the observation of D0 collaboration, the LHCb, CMS, CDF, and ATLAS Collaborations negated the existence of $X(5568)$~\cite{Aaltonen:2017voc,Aaboud:2018hgx,Sirunyan:2017ofq,Aaij:2016iev}. Thus, the present observation of $X_0(2900)$ and $X_1(2900)$ brings physicists' attentions back to the existence of fully open flavor states again.

Considering four different flavor quark components of $X_0(2900)$ and $X_1(2900)$, one can naturally consider these states as tetraquark candidates. In Ref.~\cite{Cheng:2020nho}, the mass spectrum of exotic tetraquark states with four different flavors is investigated by using a color-magnetic interaction model, where the masses of states with $I(J^P)=1(0^+)$ were 2607 and 3129 MeV, while those with $I(J^P)=0(0^+)$ were 2320 and 2850 MeV. After the observation of $X_0(2900)$ and $X_1(2900)$, the authors in Refs.~\cite{Karliner:2020vsi,Zhang:2020oze} indicated that the $X_0(2900)$ can be an isosinglet compact tetraquark state, while the estimations in Ref.~\cite{He:2020jna} indicated that the $X_0(2900)$ should be a radial excited tetraquark with $J^P=0^+$. As for $X_0(2900)$, the investigations in Refs.~\cite{Chen:2020aos, He:2020jna} support that the $X_1(2900)$ can be assigned as a $P-$wave compact diquark-antidiquark tetraquark state. However, the calculaions in an extended relativized quark model indicate that the predicted mass of $0^+$ $ud\bar{s}\bar{c}$ are different from that of the $X_0(2900)$
,  which disfavors the assignment of the $X_0(2900)$ as a compact tetraquark~\cite{Lu:2020qmp}. 

It should be noticed that in the vicinity of $2900$ MeV, there are abundant thresholds of a charmed and a strange mesons, such as $K^\ast D^\ast $, $ KD_1$, $ KD_0$. In Refs.~\cite{Guo:2011dd,Molina:2010tx}, the possible molecular states composed of (anti-)charmed and strange mesons have been investigated. Considering the $J^P$ quantum numbers of $X_0(2900)$ and $X_1(2900)$, the former one can be resulted from the $ K^\ast \bar{D}^\ast$ interaction, while the latter one can be resulted from the $K\bar{D}_1 $ interaction. In Ref.~\cite{Liu:2020orv}, the structure corresponding to $X_0(2900)$ and $X_1(2900)$ can be interpreted as the triangle singularity. While in Ref.~\cite{Liu:2020nil}, the estimation in one-boson exchange model indicated that the interaction of $ K^\ast \bar{D}^\ast$ are strong enough to form a molecular state, thus,  $X_0(2900)$ can be interpreted as a $ K^\ast \bar{D}^\ast$ molecular state and such an interpretation is also supported by the estimations in Refs.~\cite{Chen:2020aos,Hu:2020mxp}.

  Along the way of molecular interpretations, we construct the one-boson-exchange potential of $K^\ast \bar{D}^\ast$ and $ K\bar{D}_1$ interactions.  The scattering amplitude can be obtained with the help of the quasipotential Bethe-Salpeter equation (qBSE) from the interaction potentials, and the poles of the scattering amplitudes  are searched in complex energy plane. In the current work, both  bound and virtual states will be considered in the calculation to discuss the relation between experimentally observed states  $X_0(2900)/X_1(2900)$ and the $K^\ast \bar{D}^\ast/K\bar{D}_1$ interactions.

This work is organized as follows. We present the formalism used in the present estimation in the following section. The numerical results and related discussions are given in section \ref{Sec: results} and the last section is devoted to a short summary.

\section{Formalism}\label{Sec: Formalism}

In the current work, we will consider two interactions, $K^*\bar{D}^*$ and $K\bar{D}_1$ interactions. The possible isospins of the states composed by $K^*\bar{D}^*$ and $K\bar{D}_1$ could be 0 and 1, and the corresponding flavor functions are 
\begin{eqnarray}
| K^\ast \bar{D}^{\ast}, I=0 \rangle & =&\frac{1}{\sqrt{2}} \left[ K^{\ast+} D^{\ast -} - K^{\ast 0} \bar{D}^{\ast 0}\right],\nonumber\\
| K^\ast \bar{D}^{\ast}, I=1 \rangle &=&\frac{1}{\sqrt{2}} \left[ K^{\ast+} D^{\ast -} + K^{\ast 0} \bar{D}^{\ast 0}\right] , 	\nonumber\\
| K \bar{D}_1, I=0 \rangle & =&\frac{1}{\sqrt{2}} \left[ K^{+} D_1^{-} - K^{0} \bar{D}_1^{ 0}\right],\nonumber\\
| K \bar{D}_1, I=1 \rangle & =&\frac{1}{\sqrt{2}} \left[ K^{+} D_1^{-} + K^{0} \bar{D}_1^{ 0}\right],
\end{eqnarray}
respectively.

 In the one-boson-exchange model, the $K^*$ meson and $\bar{D}^*$ meson interact by exchanging $\pi$, $\eta$, $\rho$,  and $\omega$ mesons. For the $K\bar{D}_1$ interaction, the $\pi$ and $\eta$ exchanges are forbidden, only vector exchanges are allowed. To describe the interaction, we need the effective Lagrangians at two vertices. For the charmed meson part,  the effective Lagrangians can be written with the help of  heavy quark and chiral symmetries as~\cite{Cheng:1992xi,Yan:1992gz,Wise:1992hn,Casalbuoni:1996pg,Ding:2008gr},
\begin{align}\label{eq:lag-p-exch}
  \mathcal{L}_{P^*P^*\mathbb{P}}   &=
  \frac{2g}{f_\pi}\epsilon_{\mu\nu\alpha\beta} \left(P^{*\mu}_b P^{*\nu\dag}_a+\tilde{P}^{*\mu}_a\tilde{P}^{*\nu\dag}_b\right)v^\alpha \partial^\beta\mathbb{P}_{ba},\nonumber\\
  \mathcal{L}_{P^*P^*\mathbb{V}} &= \sqrt{2}\beta g_V
  \left( P^*_b\cdot P_a^{*\dag} -\tilde{P}^*_a\cdot\tilde{P}_b^{*\dag}\right)~v\cdot\mathbb{V}_{ba}\nonumber\\
&-i2\sqrt{2}\lambda
  g_V \left(P^{*\mu}_b P^{*\nu\dag}_a-\tilde{P}^{*\mu}_a\tilde{P}^{*\nu\dag}_b\right)(\partial_\mu\mathbb{V}_\nu-\partial_\nu\mathbb{V}_\mu)_{ba},
  \nonumber\\
  \mathcal{L}_{P_1P_1\mathbb{V}}
  &= \sqrt{2}\beta_2 g_V \left({P}_{1b}\cdot{P}^{\dag}_{1a}-\tilde{P}_{1a}\cdot \tilde{P}^{\dag}_{1b}\right)~v\cdot \mathbb{V}_{ba}\nonumber\\
  &+\frac{5\sqrt{2}i\lambda_2 g_V}{3}\left({P}^\mu_{1b}{P}^{\nu\dag}_{1a}-\tilde{P}^\mu_{1a}\tilde{P}^{\nu\dag}_{1b}\right)(\partial _\mu\mathbb{V}_{\nu}-\partial_\nu\mathbb{V}_\mu)_{ba},\ \ \ \ \
\end{align}
where  the velocity $v$ should be replaced by $i\overleftrightarrow{\partial}/\sqrt{m_im_f}$ with the $m_{i,f}$ being the mass of the initial or final heavy meson. The $\mathbb
P$ and $\mathbb V$ are the pseudoscalar and vector matrices as
\begin{equation}
    {\mathbb P}=\left(\begin{array}{ccc}
        \frac{\sqrt{3}\pi^0+\eta}{\sqrt{6}}&\pi^+&K^+\\
        \pi^-&\frac{-\sqrt{3}\pi^0+\eta}{\sqrt{6}}&K^0\\
        K^-&\bar{K}^0&-\frac{2\eta}{\sqrt{6}}
\end{array}\right),
\mathbb{V}=\left(\begin{array}{ccc}
\frac{\rho^0+\omega}{\sqrt{2}}&\rho^{+}&K^{*+}\\
\rho^{-}&\frac{-\rho^{0}+\omega}{\sqrt{2}}&K^{*0}\\
K^{*-}&\bar{K}^{*0}&\phi
\end{array}\right),\label{MPV}
\end{equation}
which correspond to $(\bar{D}^0,D^-,D_s^-)$.  
The coupling constants  have been determined in the literature with the heavy quark symmetry and available experimental data,  i. e., $g=0.59$, $\beta=0.9$, $\lambda=0.56$, $\beta_2=1.1$, $\lambda_2=-0.6$,  with $g_V=5.9$ and $f_\pi=0.132$ GeV~\cite{Chen:2019asm,Liu:2011xc,Isola:2003fh,Falk:1992cx,Dong:2019ofp,He:2019csk}.

To describe the couplings of the $K^{(*)}$ meson with exchanged  pseudoscalar and/or  vector mesons,  the effective Lagrangians are adopted as
\begin{align}
{\cal L}_{KKV}&=-ig_{KKV} ~KV^\mu \partial_\mu K+{\rm H.c.},\nonumber\\
	{\cal L}_{K^*K^*V}&=i\frac{g_{K^*K^*V}}{2}( K^{*\mu\dag}{V}_{\mu\nu}K^{*\nu}+K^{*\mu\nu\dag}{V}_{\mu}K^{*\nu}+K^{*\mu\dag}{V}_{\nu}K^{*\nu\mu}),\nonumber\\
	{\cal L}_{K^*K^*P}&=g_{K^*K^*P}\epsilon^{\mu\nu\sigma\tau}\partial^\mu K^{*\nu}
	\partial_\sigma P K^{*\tau}+{\rm H.c.},
\end{align}
where $K^{*\mu\nu}=\partial^\mu K^{*\nu}-\partial^\nu K^{*\mu}$. The flavor structures are  $K^{*\dag}{\bm {A}}\cdot {\bm \tau} K^*$ for an isovector $A$ ($=\pi$ or $\rho$) meson,  and $K^{*\dag}  K^* B$  for an isoscalar $B$ ($=\eta$, $\omega$) meson. 
With the help of the  SU(3) symmetry, the coupling constants can be obtained  from the $\rho\rho\rho$ and $\rho\omega\pi$ couplings.  The $g_{\rho\rho\rho}$ was suggested equivalent to $g_{\pi\pi\rho}=6.2$, and $g_{\omega\pi\rho}=11.2$ GeV$^{-1}$~\cite{Matsuyama:2006rp, Bando:1987br,Janssen:1994uf}.  The SU(3) symmetry suggests $g_{K^*K^*\rho}=g_{K^*K^*\omega}=g_{\rho\rho\rho}/(2\alpha)$, and $g_{K^*K^*\pi}=g_{K^*K^*\eta}/[-\sqrt{1/3}(1-4\alpha)]=g_{\omega\rho\pi}/(2\alpha)$ with $\alpha=1$~\cite{Ronchen:2012eg, He:2018plt, He:2017aps,He:2015yva}.

In fact, the above vertices has been applied to study many XYZ particles and hidden-strange molecular states~\cite{He:2018plt, He:2017aps,He:2015yva, He:2015cea,He:2019ify,He:2019csk,He:2019rva}. Hence, in the current work, we only need to reconstruct the vertices to the potential considered here as
\begin{equation}%
{\cal V}_{\mathbb{P}}=I_{\mathbb{P}}\Gamma_1\Gamma_2 P_{\mathbb{P}}f_\mathbb{P}^2(q^2),\ \ 
{\cal V}_{\mathbb{V}}=I_{\mathbb{V}}\Gamma_{1\mu}\Gamma_{2\nu}  P^{\mu\nu}_{\mathbb{V}}f_\mathbb{V}^2(q^2),\label{V}
\end{equation}
where the propagators are defined as usual as
\begin{equation}%
P_{\mathbb{P}}= \frac{i}{q^2-m_{\mathbb{P}}^2},\ \
P^{\mu\nu}_\mathbb{V}=i\frac{-g^{\mu\nu}+q^\mu q^\nu/m^2_{\mathbb{V}}}{q^2-m_\mathbb{V}^2},
\end{equation}
and we adopt a form factor $f_e(q^2)$  to compensate the off-shell effect of exchanged meson as $f_e(q^2)=e^{-(m_e^2-q^2)^2/\Lambda_e^2}$
with $m_e$ being the $m_{\mathbb{P},\mathbb{V}}$ and $q$ being the momentum of the exchanged  meson. Such treatment also reflects the non-pointlike nature of the constituent mesons. The cutoff is rewritten as a form of $\Lambda_e=m+\alpha_e~0.22$ GeV. The flavor factors $I_e$ for certain meson exchange and total isospin are presented in Table~\ref{flavor factor}.
\renewcommand\tabcolsep{0.43cm}
\renewcommand{\arraystretch}{1.5}
\begin{table}[h!]
\caption{The flavor factors $I_e$ for certain meson exchange  and total isospin.  The  $\pi$ and $\eta$ exchanges are forbidden for $K\bar{D}_1$ interaction.  \label{flavor factor}}
\begin{tabular}{c|ccccc}\bottomrule[2pt]
& $I_\pi$& $I_\eta$ &$I_\rho $ & $I_\omega $ \\\hline
 $I=0$&$-3\sqrt{2}/2$ &${1}/{\sqrt{6}}$  & $-3\sqrt{2}/2$&${1}{\sqrt{2}}$\\
 $I=1$&$\sqrt{2}/2$ &${1}/{\sqrt{6}}$ &$\sqrt{2}/2$&${1}/{\sqrt{2}}$\\
\toprule[2pt]
\end{tabular}

\end{table}

With the potential, the scattering amplitude can be obtained with the qBSE~\cite{Gross:2010qm,He:2012zd,He:2011ed}. The qBSE with fixed spin-parity $J^P$ is written as~\cite{He:2015cca,He:2015mja,He:2017aps},
\begin{eqnarray}
i{\cal M}^{J^P}_{\lambda'\lambda}({\rm p}',{\rm p})
&=&i{\cal V}^{J^P}_{\lambda',\lambda}({\rm p}',{\rm
p})+\sum_{\lambda''}\int\frac{{\rm
p}''^2d{\rm p}''}{(2\pi)^3}\nonumber\\
&\cdot&
i{\cal V}^{J^P}_{\lambda'\lambda''}({\rm p}',{\rm p}'')
G_0({\rm p}'')i{\cal M}^{J^P}_{\lambda''\lambda}({\rm p}'',{\rm
p}),\quad\quad \label{Eq: BS_PWA}
\end{eqnarray}
where the sum extends only over nonnegative helicity $\lambda''$.  
The $G_0({\rm p}'')$ is reduced from 4-dimensional  propagator by the spectator approximation, and  in the center-of-mass frame with $P=(W,{\bm 0})$ it reads,
\begin{align}
	G_0({ \rm p}'')
&=\frac{1}{2E_h({\rm p''})[(W-E_h({\rm
p}''))^2-E_l^{2}({\rm p}'')]}.\label{G0}
\end{align}
Here, as required by the spectator approximation, the heavier  meson ($h=\bar{D}^*,\bar{D}_1$) is on shell, which satisfies  $p''^0_h=E_{h}({\rm p}'')=\sqrt{
m_{h}^{~2}+\rm p''^2}$. The $p''^0_l$ for the lighter meson ($l=K^*, K$) is then $W-E_{h}({\rm p}'')$. A definition ${\rm p}=|{\bm p}|$ will be adopted here. The partial-wave potential is defined with the potential of the interaction obtained in the above as
\begin{eqnarray}
{\cal V}_{\lambda'\lambda}^{J^P}({\rm p}',{\rm p})
&=&2\pi\int d\cos\theta
~[d^{J}_{\lambda\lambda'}(\theta)
{\cal V}_{\lambda'\lambda}({\bm p}',{\bm p})\nonumber\\
&+&\eta d^{J}_{-\lambda\lambda'}(\theta)
{\cal V}_{\lambda'-\lambda}({\bm p}',{\bm p})],
\end{eqnarray}
where $\eta=PP_1P_2(-1)^{J-J_1-J_2}$ with $P$ and $J$ being parity and spin for system, $K^*/K$ meson or $\bar{D}^*/\bar{D}_1$ meson. The initial and final relative momenta are chosen as ${\bm p}=(0,0,{\rm p})$  and ${\bm p}'=({\rm p}'\sin\theta,0,{\rm p}'\cos\theta)$. The $d^J_{\lambda\lambda'}(\theta)$ is the Wigner d-matrix. 
In the qBSE approach,  a form factor will be introduced into the propagator to reflect the off-shell effect as an exponential regularization,
$
G_{0}(p)\rightarrow G_{0}(p)[e^{-(k^{2}_{1}-m^{2}_{1})^{2}/\Lambda^{4}_{r}}]^{2},
$
where the $k_{1}$ and  $m_{1}$ are the momentum and the mass of the strange meson.   The cutoff $\Lambda_{r}$ is also parameterized as in the $\Lambda_{e}$ case.  The $\alpha_e$ and $\alpha_r$ play  analogous roles in the calculation of the binding energy. Hence, we take these two parameters as a parameter $\alpha$ for simplification~\cite{He:2019csk}.  Such parameter is also used to absorb the uncertainties of our model, such as the inaccuracy of heavy quark and SU(3) symmetries in the Lagrangians.

\section{Numerical results and discussion}\label{Sec: results}

With the scattering amplitude obtained above, the pole can be searched for in the complex energy plane.  The bound state corresponds to a pole at the real axis  under threshold in the first Riemann surface. If the attraction becomes weaker, the pole will move to the real axis  under threshold in the second Riemann surface, which corresponds to a virtual state~\cite{Taylor}. In the current work, we will consider both bound and virtual states from the $K^*\bar{D}^*$ and $K\bar{D}_1$ interactions. 

\subsection{States from $K^*\bar{D}^*$  interaction}\label{Sec: KADA}

\begin{figure}[h!]
\begin{center}
\includegraphics[width=90mm]{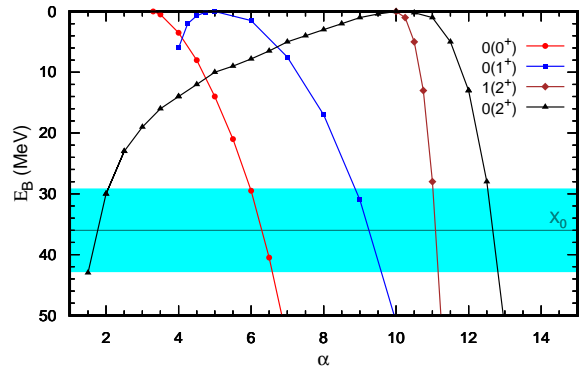}
\end{center}
\caption{The binding energy $E_B$ of the bound or virtual states from the $K^*\bar{D}^*$ interaction with the variation of $\alpha$. Here the  $E_B=M_{th}-W$ with the $M_{th}$ and $W$ being the threshold and mass of the state. The circle, square, diamond, and triangle are for the states with $I(J^P)=0(0^+)$, $0(1^+)$, $1(2^+)$, and $0(2^+)$, respectively. The lines with cyan bar are for experimental mass and uncertainties of  $X_0(2900)$ state, respectively. 
}
\label{Fig: KADA}
\end{figure}

In the current work, we will consider six states from the $K^*\bar{D}^*$  interaction with isospin $I=(0, 1)$, spin $J=(0, 1, 2)$, and parity $P=+$ which can be obtained in S wave.  In our model, the only free parameter is the $\alpha$ in cutoff.  Usually, small value of $\alpha$ should be chosen.  For a cutoff $\Lambda$ smaller than 3 GeV, the $\alpha$ should be smaller than $10$. In the following, we present the results with $\alpha$  value in a larger range  from 1 to 20 for discussion. We would like to remind in advance that the results with very large $\alpha$ are unreliable because it corresponds to a very small radius of the constituent hadrons.  The results for the states from the $K^*\bar{D}^*$  interaction are presented  and compared with experimentally observed $X_0(2900)$ in Fig.~\ref{Fig: KADA} (here we call the deviation between the pole of a virtual state and threshold as binding energy also).

 Among the six states considered in the current work, four bound states can be produced from the $K^*\bar{D}^*$ interaction in the large range of $\alpha$ considered here.  The bound states with $I(J^P)=0(0^+)$ and $0(1^+)$ appear at small $\alpha$, about 4, and two bound states with $2^+$ are found at $\alpha$ larger than 10.  Usually, larger cutoff  corresponds  to stronger interaction, which leads to larger binding energy for a bound state. One can find that the binding energies of the four bound states increase with the increase of the $\alpha$ value. 
 
 Here, we also consider the possible virtual state from the interaction.  Different from bound state,  virtual state leaves the threshold further with the decreasing of  $\alpha$ and weakening of  attraction.  The bound state with $I(J^P)=0(2^+)$ appears at $\alpha$ about 10, and the energy increases rapidly with the increase of the $\alpha$ value.   However, if we reduce the $\alpha$ value, a pole can be found at second Riemann surface, and leaves the threshold with the decrease of $\alpha$ value.  The pole  moves to a position about 40 MeV below the threshold at an $\alpha$ about 2, and disappears there.  No virtual state can be found for the case with $0(0^+)$ and $1(2^+)$ if we reduce the  $\alpha$ value.  For $0(1^+)$ case, virtual state is also found, but disappears very rapidly with the decrease of  $\alpha$ value.  
 
Among the four bound states produced from the  $K^*\bar{D}^*$ interaction, two bound states with $0(0^+)$ and $0(1^+)$ require small  $\alpha$ value. For the $0(2^+)$ state, only virtual state can be produced with small $\alpha$ value. Since the $X_{0}(2900)$ and $X_1(2900)$ were observed in the $K^+D^-$ channel, allowed quantum numbers of are $0^+$ and $1^-$.  Hence, the current results support the assignment of  the $X_0(2900)$ observed at LHCb as a $0(0^+)$ state from  the  $K^*\bar{D}^*$ interaction.  Under such assignment, a bound state with $0(1^+)$ and a virtual state with $0(2^+)$ are also predicted in our model.

\begin{figure}[htb ]
\begin{center}
\includegraphics[width=90mm]{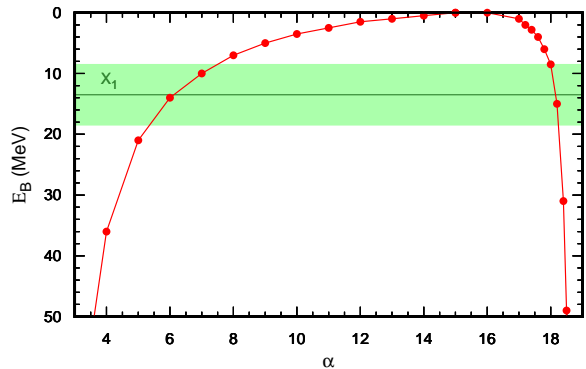}
\end{center}
\caption{The binding energy $E_B$ of the bound or virtual state from the $K\bar{D}_1$ interaction with the variation of $\alpha$. The circle is for the states with $I(J^P)=0(1^-)$. The lines with lightgreen bar are for experimental mass and uncertainties of  $X_1(2900)$  state, respectively.  Other conventions are the same as in Fig. \ref{Fig: KADA}.
}
\label{Fig: KD1}
\end{figure}

\subsection{States from $K\bar{D}_1$  interaction}\label{Sec: w}

The $X_1(2900)$ state can not be reproduced from  the  $K^*\bar{D}^*$ interaction in S wave. Here we consider another system with a threshold close to the mass of $X_1(2900)$,  the $K\bar{D}_1$  interaction. We will consider two states from the $K\bar{D}_1$  interaction with $I=(0, 1)$ and $J^P=1^-$, which can be obtained in S wave.  The results are presented in Fig.~\ref{Fig: KD1}.

Among these two states,  only the isoscalar interaction is attractive. However, the bound state with $0(1^-)$ appears at a very larger $\alpha$ value, about 16, which corresponds to a large cutoff $\Lambda$ about 4 GeV.  It is unreliable to assign the  $X_1(2900)$ as a bound state. As the $0(2^+)$ state of the $K^*D^*$ interaction, if we decrease the $\alpha$ value, a virtual state with $0(1^-)$ from the $K\bar{D}_1$ interaction can be found in a large range of the $\alpha$ form about 4 to 16.  Such state can be related to the experimentally observed $X_0(2900)$.

\section{Summary}\label{Sec: summary}

In the current work, inspired by the newly observed $X_{0,1}(2900)$ at LHCb, the $K^*\bar{D}^*$ and $K\bar{D}_1$ interactions, which have thresholds about 2900 MeV,  are studied in the qBSE approach. The bound and virtual states from the interaction are searched for as poles in the complex energy plane of the scattering amplitude, which is obtained from the one-boson-exchange potential. A bound state with $0(0^+)$ is produced  from the $K^*\bar{D}^*$ interaction. The radius $R$ of the bound state state can be estimated as  $R\sim1/\sqrt{2\mu E_B}$ with $\mu$ and $E_B$ being the reduced mass and binding energy~\cite{Guo:2017jvc}. The experimental binding energy, about 35 MeV, leads to a radius about 1~fm of the $K^*\bar{D}^*$ bound state.  Considering the constituent mesons have radii about 0.5~fm, it supports the assignment of  $X_0(2900)$ as a $K^*\bar{D}^*$ molecular state.   A virtual state with $0(1^-)$ is also produced from $K\bar{D}_1$ interaction with reasonable parameter.  These two states can decay into  the $K^+D^-$ channel in S and P waves, so can be related the $X_{0}(2900)$  and $X_1(2900)$ observed at LHCb, respectively. Besides these two states, a bound  state with $0(1^+)$ and a virtual state with $0(2^+)$ are produced from the $K^*\bar{D}^*$ interaction with a small $\alpha$ value.  

\section*{Acknowledgments}

This project is supported by the National Natural Science
Foundation of China (Grants No. 11675228, and No. 11775050), and the Fundamental Research Funds for the Central Universities.

\end{document}